\documentclass[prl,twocolumn,
nofootinbib,preprintnumbers]{revtex4}

\usepackage{times}
\usepackage{natbib}
\usepackage{amssymb,amsbsy,amsmath,amsfonts}
\usepackage{mathrsfs}
\usepackage{graphicx}
\usepackage{epsf,epsfig,float,latexsym,amsthm,fancyhdr,rotating}
\usepackage{graphics,psfrag,longtable}
\usepackage{slashed}
\usepackage{esint}
\usepackage{nicefrac}

\DeclareMathAlphabet{\mathantt}{OML}{antt}{l}{it}
\DeclareMathAlphabet{\mathpzc}{OT1}{pzc}{m}{n}

\def\beq{\begin{equation}}
\def\eeq{\end{equation}}
\def\bea{\begin{eqnarray}}
\def\eea{\end{eqnarray}}
\def\beqa{\begin{equation}\begin{array}{l}}
\def\eeqa{\end{array}\end{equation}}

\def\barr{\left(\begin{array}{c}}
\def\earr{\end{array}\right)}
\def\bmat{\left(\begin{array}{cc}}
\def\emat{\end{array}\right)}

\begin{document}
\title {Lepton universality test in the photoproduction of $e^- e^+$ versus $\mu^-\mu^+$ pairs on a proton target}

\author{Vladyslav Pauk}
\author{Marc Vanderhaeghen}

\affiliation{
Institut f\"ur Kernphysik, Cluster of Excellence PRISMA,  Johannes Gutenberg-Universit\"at, D-55099 Mainz, Germany}

\begin{abstract}
In view of the significantly different proton charge radius extracted from muonic hydrogen Lamb shift measurements as compared to electronic hydrogen spectroscopy or electron scattering experiments, we study in this work the photoproduction of a lepton pair on a proton target in the limit of very small momentum transfer as a way to provide a test of the lepton universality when extracting the proton charge form factor.   By detecting the recoiling proton in the 
$\gamma p \to l^- l^+ p$ reaction, we show that a measurement of a ratio of $e^-e^+ + \mu^-\mu^+$ over $e^-e^+$ cross sections  with an absolute precision  of $7 \times 10^{-4}$, would allow for a test to distinguish, at the $3 \sigma$ level, between the two different proton charge radii currently extracted from muonic  and electronic observables.   
\end{abstract}
\pacs{13.40.Gp, 13.60.Fz, 14.20.Dh, 14.60.Ef}
\date{\today}
\maketitle

Recent extractions of the proton charge radius $R_E$ from muonic hydrogen Lamb shift measurements~\cite{Pohl:2010zza, Antognini:1900ns}  are in strong contradiction, by around 7 standard deviations, with the values obtained from energy level shifts in electronic hydrogen~\cite{Mohr:2012tt} or from electron-proton elastic scattering~\cite{Bernauer:2010wm, Bernauer:2013tpr}~:
\begin{eqnarray}
\mu p~\text{\cite{Antognini:1900ns}} : \quad &&R_E = 0.8409\, (4)~{\mathrm {fm}} \, ,\label{eq:radiusmu} \\
e p~\text{\cite{Mohr:2012tt}}: \quad &&R_E = 0.8775 \, (51)~{\mathrm {fm}} \, . \label{eq:radiusel}
\end{eqnarray}
This so-called ''proton radius puzzle" has triggered a large activity and is  the subject of intense debate, see e.g.~\cite{Pohl:2013yb, Bernauer:2014cwa, Carlson:2015jba} for recent reviews, and references therein. 
Lepton universality requires the same radius to enter the  
electronic and muonic observables. If the different $R_E$ extractions cannot be explained by overlooked corrections, it would point to a violation of electron-muon universality. Several scenarios of new, beyond 
the Standard Model, physics have been proposed by invoking new particles which couple to muons and protons, but much weaker to electrons, see e.g.~\cite{TuckerSmith:2010ra, Barger:2010aj, Batell:2011qq, Barger:2011mt, Carlson:2012pc}.  Such models would also lead to large loop corrections to the muon's anomalous magnetic moment, $(g-2)_\mu$, which presently displays a $3 \sigma$ deviation between experiment and its Standard Model prediction, see e.g.~\cite{Blum:2013xva}. Explaining both the $(g-2)_\mu$ discrepancy and the proton radius puzzle by new particles coupling mainly to muons seems an attractive perspective. It does however require a significant fine-tuning, especially for larger values of the conjectured new particle masses~\cite{Carlson:2015jba}. 
To test the electron-muon universality, it has been proposed by the 
MUon proton Scattering Experiment (MUSE) \cite{Gilman:2013eiv} to make a simultaneous measurement of both $\mu p$ and $e p$ elastic scattering, extracting the proton charge form factor from both measurements. Besides the plans to measure $\mu p$ elastic scattering, several new experiments are underway to extend $e p$ scattering to lower momentum transfer values, down to $10^{-4}$~GeV$^2$, and to cross-check its 
systematics~\cite{Mihovilovic:2014aya, Gasparian:2014rna}.  All of these 
tests require absolute cross section measurements, with a required precision on each absolute cross section at the level of 1 \% or better. 

In order to reach in $\mu$ scattering experiments the precision on $R_E$ 
comparable with $e^-$ scattering experiments, indicated in Eq.~(\ref{eq:radiusel}), 
we propose in this work a new experimental avenue through a relative cross section measurement of the photo-production of $e^- e^+$ versus $\mu^- \mu^+$ pairs on a proton target. Besides being a well studied process~\cite{Motz:1969ti}, the photo-production of a lepton pair has the advantage that one produces $e^- e^+$ and $\mu^- \mu^+$ final states with the same beam, and thus the overall normalization uncertainty drops out of their ratio.  The analysis presented in this work shows that through a detection of the recoiling proton in the $\gamma p \to l^- l^+ p$ reaction, a measurement of the ratio of the $e^- e^+$ cross section below $\mu^- \mu^+$ threshold versus the $e^- e^+ + \mu^- \mu^+$ cross section sum above $\mu^- \mu^+$ threshold with an absolute precision of around $7 \times 10^{-4}$ will allow to distinguish, at the $3 \sigma$ level, between the proton $R_E$ extractions from muonic and electronic observables. Furthermore, we show that the linear photon polarization asymmetry has a discriminatory power between the $e^- e^+$ and $\mu^- \mu^+$ channels in the $\mu^- \mu^+$ threshold region. 
\begin{figure}[h]
\centering
\includegraphics[width=0.45\textwidth]{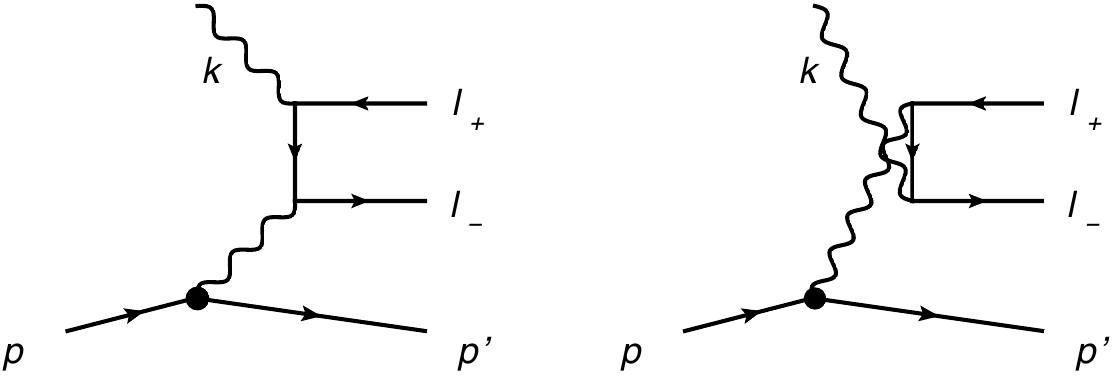}
\caption{Bethe-Heitler mechanism to the $\gamma p \to l^-l^+ p$ process, where the four-momenta of the external particles are: 
$k$ for the photon, $p (p^\prime)$ for initial (final) protons, and $l_-$, $l_+$ for the lepton pair.} 
\label{fig1}
\end{figure}

We will consider the lepton pair production on a proton target, 
$\gamma p \to l^- l^+ p$, in the limit of 
very small momentum transfer, defined as $t = (p - p^\prime)^2$, with 
four-momenta as indicated on Fig.~\ref{fig1}. Furthermore, we will use in the following the Mandelstam invariant $s = (k + p)^2 = M^2 + 2 M E_\gamma$, with $M$ the proton mass and $E_\gamma$ the photon {\it lab} energy, as well as the squared invariant mass of the lepton pair, defined as 
$M_{ll}^2 \equiv (l_- + l_+)^2$. 
In the limit of small $-t$, the Bethe-Heitler (BH) mechanism, shown in Fig.~\ref{fig1} totally dominates the cross section of the $\gamma p \to l^- l^+ p$ reaction.  
As the momentum transfer $t$ is the argument appearing in the form factor (FF) in the BH process, a measurement of the cross section in this kinematic regime will allow to access the proton electric FF $G_{Ep}$ at small spacelike momentum transfer, with a very small contribution of the proton magnetic FF $G_{Mp}$. As the differential cross section for the BH process is strongly peaked for leptons emitted in the incoming photon direction, and as we aim to maximize the BH contribution in this work in order to access $G_{Ep}$, 
we will study the $\gamma p \to l^- l^+ p$  process when (only) detecting the recoiling proton's momentum and angle, thus effectively integrating over the large lepton peak regions.  
The {\it lab} momentum of the proton is in one-to-one relation with the momentum transfer $t$: 
$|\vec p^{\, \prime}|^{lab} = 2 M \sqrt{\tau (1 + \tau)}$, 
with $\tau \equiv -t / (4 M^2)$. Furthermore, for a fixed value of $t$, 
the recoiling proton {\it lab} 
angle $\Theta_p^{lab}$ is expressed in terms of invariants as~:
\begin{eqnarray}
\cos \Theta_p^{lab} = \frac{M_{ll}^2 + 2 (s + M^2) \tau}{2 (s - M^2) 
\sqrt{\tau (1 + \tau)}}.
\end{eqnarray}

The differential cross section for the dominating BH process to 
the $\gamma p \to l^- l^+ p$ reaction has been studied 
in different contexts in the literature~\cite{Motz:1969ti, Berger:2001xd, Boer:2015fwa}. In this work, we will consider the cross section differential in the momentum transfer $t$ and invariant mass 
of the lepton pair $M_{ll}^2$, and integrated over the lepton angles, which corresponds with detecting the recoiling proton only.  This cross section can be written as~:
\begin{eqnarray}
\frac{d \sigma^{BH}}{dt \, dM_{ll}^2} &=& \frac{\alpha^3}{(s - M^2)^2} 
\cdot \frac{4 \beta}{t^2 (M_{ll}^2 - t)^4} \cdot \frac{1}{1 + \tau} \nonumber \\
&\times& \left\{ C_E \, G_{Ep}^2  
+ C_M \, \tau \, G_{M p}^2 \right\},
\label{eq:intcross}
\end{eqnarray}
with $\alpha \equiv e^2 / 4 \pi \approx 1/137$, 
where $\beta \equiv \sqrt{1 - \frac{4 m^2}{M_{ll}^2}}$ 
is the lepton velocity in the $l^- l^+$ {\it c.m.} frame, with $m$ the lepton mass, and where the proton FFs $G_{Ep}$ and $G_{M p}$ are functions of $t$.  
The weighting coefficients multiplying the FFs in Eq.~(\ref{eq:intcross})
have the following general structure~:
\begin{eqnarray}
C_{E, M} = C_{E, M}^{(1)} + C_{E, M}^{(2)} 
\frac{1}{\beta} \ln \left( \frac{1 + \beta}{1 - \beta} \right),  
\label{eq:CEM}
\end{eqnarray}
where the second term expresses the large 
logarithmic enhancement in the limit of small lepton mass in the BH process. 
The coefficients $C_{E, M}^{(1)}$, and $C_{E, M}^{(2)}$ are found to be expressed through invariants as~:
\begin{widetext}
\begin{eqnarray}
C_E^{(1)} &=& t \left(s - M^2 \right) \left( s - M^2 - M_{ll}^2 + t \right) 
\left[ M_{ll}^4 + 6 M_{ll}^2 t + t^2 + 4 m^2 M_{ll}^2 \right] \nonumber \\
&+& \left(M_{ll}^2 - t \right)^2 \left[ t^2 M_{ll}^2 + M^2 (M_{ll}^2 + t )^2 
+ 4 m^2 M^2 M_{ll}^2 \right], \\
C_E^{(2)} &=& - t \left(s - M^2 \right) \left( s - M^2 - M_{ll}^2 + t \right) 
\left[ M_{ll}^4 + t^2 + 4 m^2 \left( M_{ll}^2 + 2 t - 2 m^2\right) \right] \nonumber \\
&+& \left(M_{ll}^2 - t \right)^2 \left[ - M^2 (M_{ll}^4 + t^2 ) 
+ 2 m^2 \left( - t^2 - 2 M^2 M_{ll}^2 + 4 m^2 M^2 \right) \right], \\
C_M^{(1)} &=& C_E^{(1)} - 2 M^2 (1 + \tau)   
\left(M_{ll}^2 - t \right)^2  
\left[  M_{ll}^4 + t^2 + 4 m^2 M_{ll}^2 \right], \\
C_M^{(2)} &=& C_E^{(2)} + 2 M^2 (1 + \tau)   
\left(M_{ll}^2 - t \right)^2  
\left[  M_{ll}^4 + t^2 + 4 m^2 \left( M_{ll}^2 - t - 2 m^2 \right) \right] .
\end{eqnarray}
\end{widetext}

 \begin{figure}[h]
\centering
\includegraphics[width=0.45\textwidth]{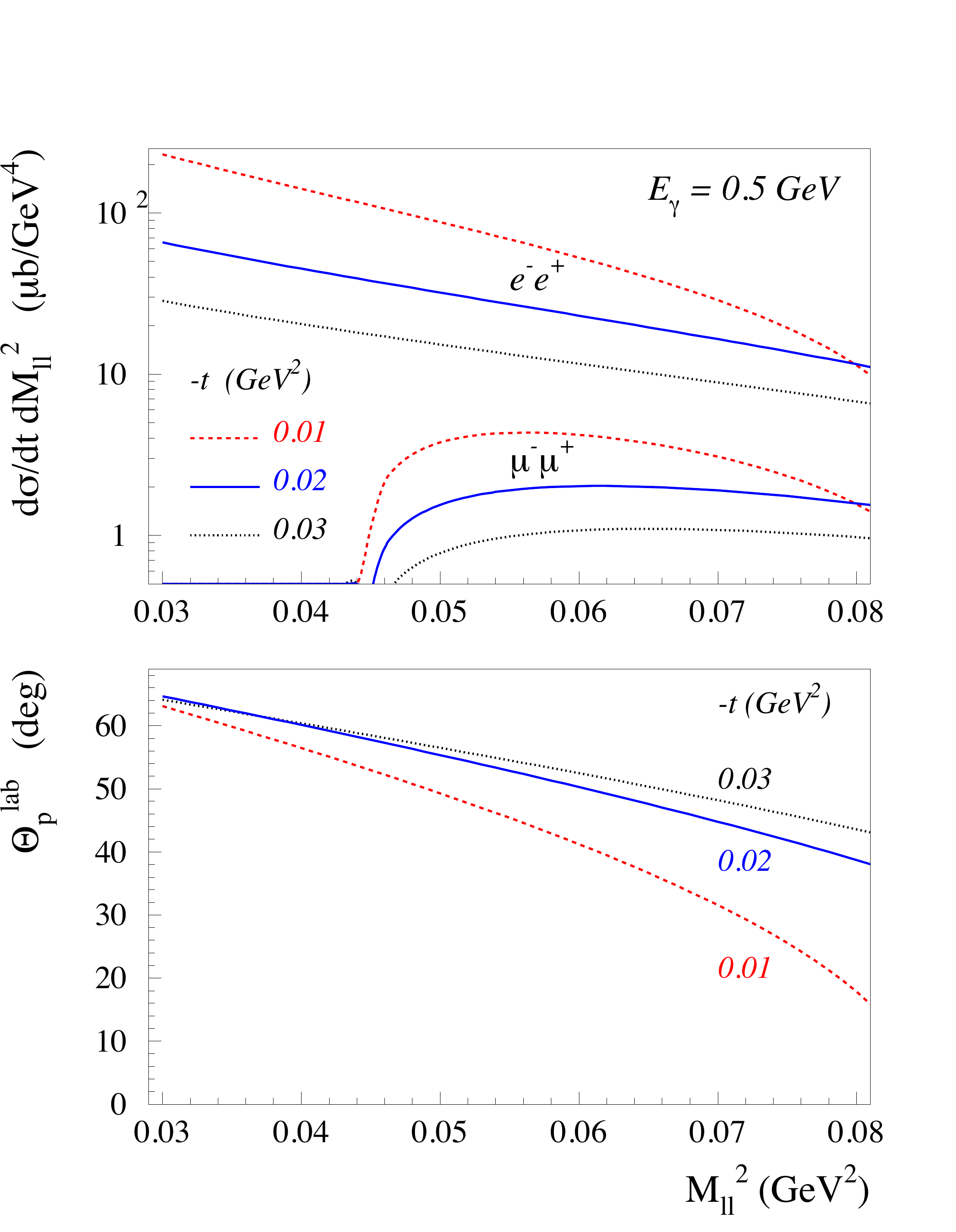}
\caption{Upper panel: comparison of the (lepton pair) invariant mass dependence of the $\gamma p \to e^-e^+ p$ process (upper three curves) vs the $\gamma p \to \mu^- \mu^+ p$ process (lower three curves) at $E_\gamma = 0.5$~GeV, and for three values of the momentum transfer $t$ as indicated. The lower panel shows the corresponding kinematic relation between the lepton pair invariant mass and the recoiling proton {\it lab} angle.} 
\label{fig2}
\end{figure}

\begin{figure}[h]
\centering
\includegraphics[width=0.45\textwidth]{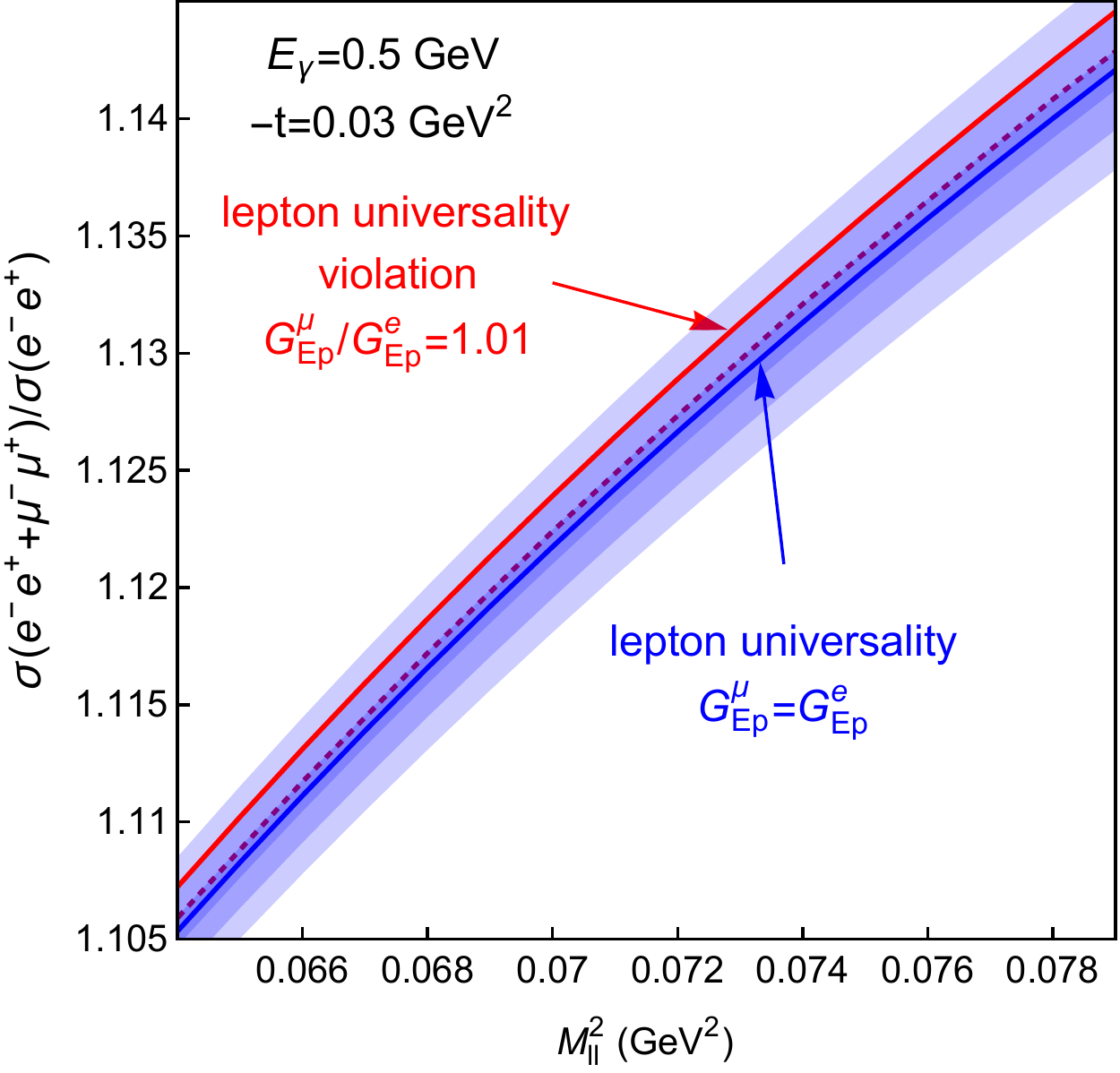}
\caption{Ratio $R_{\mu /e}$ of the 
$\gamma p \to (e^- e^+ + \mu^- \mu^+) \, p$ vs 
$\gamma p \to  (e^- e^+) \, p$ differential cross sections, according to Eq.~(\ref{eq:ratioR}). 
The lower (blue) curve corresponds with the lepton universality result, for which  $G_{Ep} ^{\mu} = G_{Ep}^{e}$. The upper (red) curve assumes a violation of lepton universality, for which $G_{Ep}^{\mu} / G_{Ep}^{e}$ = 1.01. The blue bands denote 1, 3, 5 $\sigma$ bands corresponding with a $\sigma$ of $7 \times 10^{-4}$ 
around the blue curve. The difference between the blue curve and the dotted curve  is an estimate of the physical background due to the contribution of the timelike Compton process. } 
\label{fig3}
\end{figure}

In Fig.~\ref{fig2}, we show the differential cross section 
$d \sigma / dt \, dM_{ll}^2$ for 
$\gamma p \to (l^- l^+) p$ which is accessed by measuring the recoiling proton's momentum and angle, for $E_\gamma = 0.5$~GeV and for three values of $-t$~:  
$-t = 0.01$~GeV$^2$ (corresponding with recoil proton momentum 
$|\vec p^{\, \prime}|^{lab} = 100$~MeV/c), 
$-t = 0.02$~GeV$^2$ ($|\vec p^{\, \prime}|^{lab} = 142$~MeV/c), 
and 
$-t = 0.03$~GeV$^2$ ($|\vec p^{\, \prime}|^{lab} = 174$~MeV/c). 
As $t$ is the argument entering the proton FFs, 
the values shown are chosen to cover the lower range of the high-precision elastic $ep$ scattering experiments~\cite{Bernauer:2010wm, Bernauer:2013tpr}, 
as well as the values for which the future MUSE elastic $\mu p$ scattering experiment~\cite{Gilman:2013eiv} plans to take data.  
Furthermore, the differential cross section $d \sigma / dt d M_{ll}^2$ is shown 
as function of the squared lepton invariant mass $M_{ll}^2$, which is dialed   through the recoiling  proton {\it lab} angle, as shown on the lower panel of Fig.~\ref{fig2}. 
We show the cross section in a range of $M_{ll}^2$, which is  
kinematically separated from background channels, well above the Compton and $\pi^0$ production processes on a proton, corresponding with 
sharp peaks at $M_{ll}^2 = 0$ and at $M_{ll}^2 = 0.018$~GeV$^2$ respectively, and below the threshold for $\pi \pi$ production, which starts at $M_{ll}^2 = 0.078$~GeV$^2$. One notices from Fig.~\ref{fig2} that around $M_{ll}^2 = 0.06$~GeV$^2$, the $\mu^- \mu^+$ cross section is around a factor of 10 smaller than the $e^- e^+$ cross section, and increases with increasing $M_{ll}^2$. By measuring the cross section through detecting the recoiling proton momentum and angle in the $M_{ll}^2$ window  above the $\pi^0$ peak and below $\pi \pi$ threshold, and comparing cross sections at a fixed value of $t$ above and below $\mu^- \mu^+$ thresholds, it opens the possibility for a high-precision extraction of the  cross section ratio~:
\begin{eqnarray}
R_{\mu/e} \equiv \frac{d \sigma (\mu^- \mu^+ + e^- e^+)}{d \sigma (e^- e^+)},
\label{eq:ratioR}
\end{eqnarray}
where $d \sigma$ stands for $d \sigma / dt \, dM_{ll}^2$.
The potential advantage of such a ratio measurement is that absolute normalization uncertainties to first approximation drop out. Indeed, at a fixed value of $t$, the $e^- e^+$ cross section can be fixed by measuring the cross section below $\mu^- \mu^+$ threshold, and the corresponding normalization, due to $G_{Ep}$, can be used to determine the $e^- e^+$ cross section above $\mu^- \mu^+$  threshold. A subsequent measurement of the sum of $e^- e^+ + \mu^- \mu^+$ cross sections above $\mu^- \mu^+$ threshold, then allows to extract the ratio $R_{\mu/e}$, 
which is displayed in Fig.~\ref{fig3}. 
One sees that in the kinematic range where only the $e^- e^+$ and $\mu^- \mu^+$ channels are contributing, this ratio varies between 1.10 to 1.14. 
We like to notice that corrections, notably radiative corrections, to first order also drop out of this ratio, measured at the same value of the recoiling proton momentum and angle. 
An accurate measurement of this ratio can therefore be envisaged, opening  a new perspective to perform a test of lepton universality. We have demonstrated this sensitivity in Fig.~\ref{fig3}, 
by varying the electric FF value entering the $e^- e^+$ production process, denoted by $G^e_{Ep}$, 
which we take from~\cite{Bernauer:2010wm, Bernauer:2013tpr},
versus the electric FF value entering the $\mu^- \mu^+$ production process, denoted by $G^\mu_{Ep}$. We compare the case of lepton universality $G^\mu_{Ep} = G^e_{E p}$, with a lepton universality violation scenario in which 
$G^\mu_{Ep} / G^e_{E p} = 1.01$. The latter value is motivated by the 
around 1\% larger value of $G_{Ep}^{\mu}$ at $-t = 0.03$~GeV$^2$, resulting from the proton radius as extracted from the muonic hydrogen Lamb shift of Eq.~(\ref{eq:radiusmu}), 
in comparison with the proton radius as extracted from $ep$ scattering or 
electronic hydrogen Lamb shift, given by Eq.~(\ref{eq:radiusel}). 
We like to note that more recent studies, using a more conservative treatment of systematic errors in elastic $e p$ scattering, 
extract a radius in agreement with Eq.~(\ref{eq:radiusel}), but with twice larger error~\cite{Arrington:2015ria}. 
 
We see from Fig.~\ref{fig3} that these two scenarios for the proton radius lead to a difference in the cross section ratio 
$R_{\mu / e}$ of Eq.~(\ref{eq:ratioR}) of around $2 \times 10^{-3}$.  A measurement of this ratio $R_{\mu / e}$ 
with a precision of around $7 \times 10^{-4}$ 
would therefore provide a test of such difference at the $3 \sigma$ level. 
In Fig.~\ref{fig3}, we show the $1, 3$, and $5 \sigma$ error bands corresponding with a $\sigma$ of 
$7 \times 10^{-4}$. 
Although the Compton and $\pi^0$ photoproduction backgrounds have been eliminated
by our choice of the kinematical range in $M_{ll}^2$, we also estimated the background contamination due to the $\gamma p \to (\pi^0 \gamma) p$ process~\cite{Chiang:2004pw}  
in the relevant range $M_{ll}^2 \approx 0.07$~GeV$^2$. 
We found that in order for the contamination on $R_{\mu/e}$ to be bounded by $\Delta R_{\mu/e} \leq 10^{-3}$, 
the required precision on the $\gamma p \to \pi^0 \gamma p$ cross section should be in the 10 - 30  \% range, which has already been achieved by present experiments.
Furthermore, the remaining concern is the physical background due to the indistinguishable timelike Compton scattering process, which results in the same 
final state. We estimate this timelike Compton process at the relatively low energy and momenta considered here by its Born contribution, corresponding to a nucleon intermediate state~\cite{Pasquini:2001yy}. 
The inclusion of the timelike Compton Born contribution is also shown in Fig.~\ref{fig3}. We found the effect due to the Born contribution in most kinematics around a factor of 5 smaller than the 
effect due to the 1~\% variation in the value of $G_{Ep}^\mu$, shown in Fig.~\ref{fig3}. 
Although the timelike Compton process requires further study, we like to notice that it is not totally unknown, and a moderate accuracy in the 30 \% range would be sufficient for its contamination to be bounded by $\Delta R_{\mu/e} \leq 10^{-3}$. 
We are therefore confident that the ratio of Eq.~(\ref{eq:ratioR}) is a promising observable for a lepton universality test at the 1~\% level between $G_{Ep}^\mu$ and 
$G_{Ep}^e$.  Besides providing such a test, a measurement of the 
absolute $\gamma p \to (e^- e^+) p$ cross section below $\mu^- \mu^+$ threshold will provide a useful cross-check on the extraction of $G_{Ep}$ from elastic $ep$ scattering. As the  systematics in both reactions are quite different,  
an independent measurement of $G_{Ep}$ may yield a valuable further clue to shed light on the ``proton radius puzzle".

\begin{figure}[h]
\centering
\includegraphics[width=0.41\textwidth]{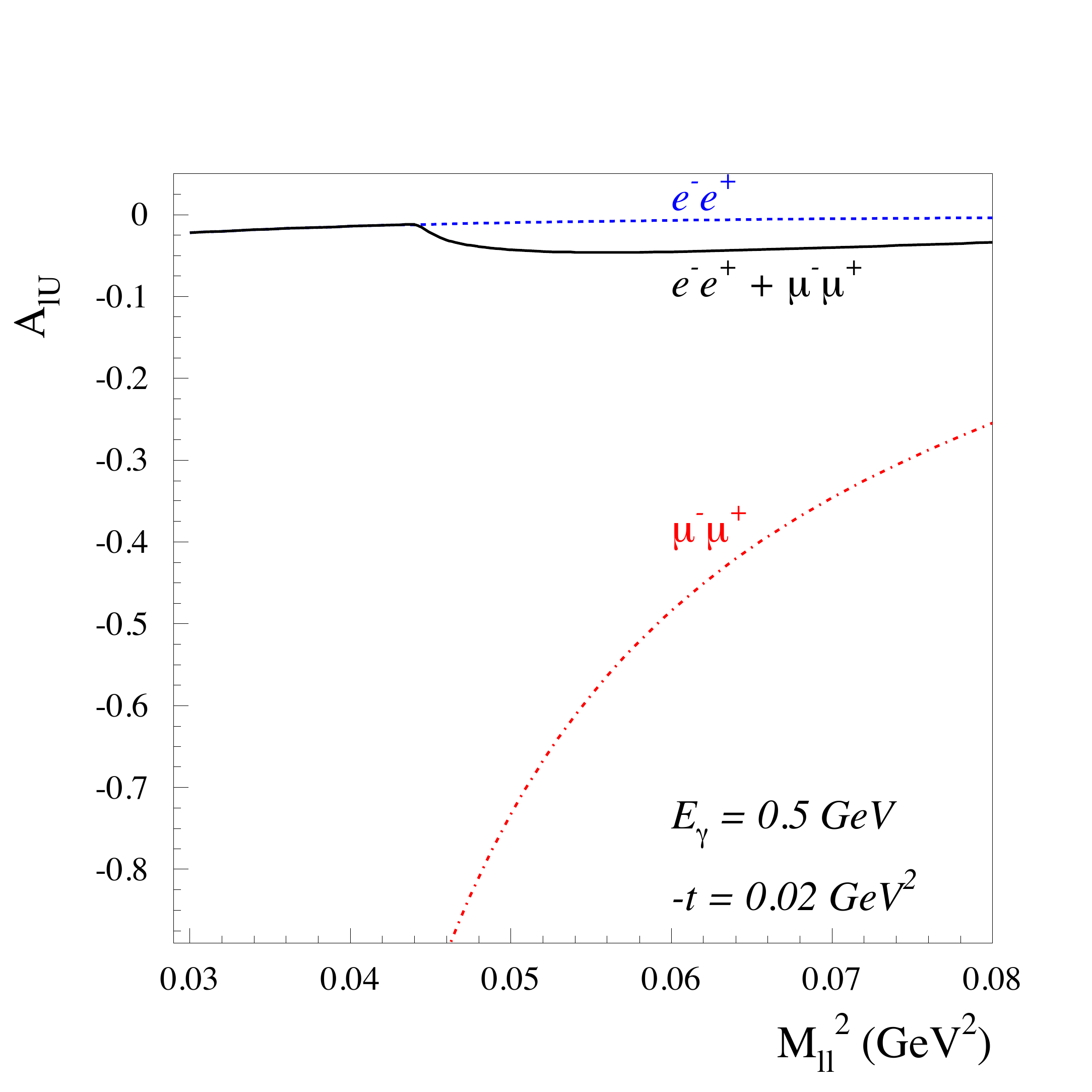}
\caption{Linear photon asymmetry $A_{lU}$ of the  
$\gamma p \to  (l^- l^+) \, p$ process. The dashed (blue)  curve corresponds with $e^-e^+$ production; the dashed-dotted (red) curve corresponds with 
$\mu^- \mu^+$ production. The solid (black) curve is the asymmetry corresponding with the sum of the $e^- e^+ + \mu^- \mu^+$ channels according to Eq.~(\ref{eq:asymmsum}).} 
\label{fig4}
\end{figure}

Besides the unpolarized cross section 
, we may also consider the sensitivity of polarization observables to distinguish between the $e^- e^+$ and $\mu^- \mu^+$ production processes. We will 
study here the case of the linear photon asymmetry defined as:
\begin{eqnarray}
A_{lU} = \frac{d \sigma_\parallel - d \sigma_\perp}{d \sigma_\parallel + d \sigma_\perp},
\end{eqnarray}
where $d \sigma_\parallel$ ( $d \sigma_\perp$ ) stands for the differential cross section for a photon with linear polarization parallel (perpendicular) to the plane spanned by the photon and recoiling proton momenta.  
When measuring the recoiling proton only, the asymmetry above 
$\mu^- \mu^+$ threshold is given by the following weighted sum of the 
asymmetries of the $e^- e^+$ and $\mu^- \mu^+$ channels: 
\begin{eqnarray}
A_{lU} (e^-e^+ + \mu^- \mu^+) &=& \frac{1}{R_{\mu/e}} \left\{ A_{lU} (e^-e^+)   \right. \nonumber \\
&+& \left.  \left( R_{\mu/e} -1 \right) \,  A_{lU} (\mu^- \mu^+)  \right\}. \hspace{0.75cm}
\label{eq:asymmsum}
\end{eqnarray}
We show in Fig.~\ref{fig4} the linear photon asymmetry in the kinematic range around $\mu^- \mu^+$ threshold. 
It is seen that the linear photon asymmetry is very small 
for the $e^- e^+$ channel. However, for the $\mu^- \mu^+$ channel the asymmetry reaches a value approaching $-1$ at $\mu^- \mu^+$ threshold and decreases in absolute value by going away from the threshold. 
Such behavior arises due to an exact cancellation,  at $\mu^- \mu^+$ threshold, 
between the analytical and lepton mass logarithmic terms in the analogous expression as Eq.~(\ref{eq:CEM}) for 
the contribution proportional to $G_{Ep}^2$ to $\sigma_\parallel$. 
In the whole $M_{ll}^2$ range of interest, the asymmetry for $\mu^- \mu^+$ production takes on large values as can be seen from Fig.~\ref{fig4}. A direct measurement of the $\mu^- \mu^+$ asymmetry may therefore give a clear tool to separate the two channels. If the lepton pair is undetected, and only the recoiling proton is measured, the measured asymmetry above $\mu^- \mu^+$ threshold is diluted by the $\mu^- \mu^+ / e^- e^+$ ratio, as given by Eq.~(\ref{eq:asymmsum}), and reaches values around -5 \% as can be seen from Fig.~\ref{fig4}.

In conclusion,  in view of the sizably different proton charge radius presently extracted from electronic and muonic observables, we proposed a new lepton universality test which meets the 
required precision goal to distinguish between them. We demonstrated that such a test is possible by comparing the photoproduction of a lepton pair on a proton, through detection of the recoiling proton. We showed that the measurement of the ratio of photoproduction cross sections of   $e^- e^+ + \mu^- \mu^+$ vs $e^- e^+$ pairs with an absolute precision of around $7 \times 10^{-4}$ will allow to distinguish, at the $3 \sigma$ level between the different proton $R_E$ extractions from muonic and electronic observables. Such an experiment can be performed at existing electron facilities such as the Mainz Mikrotron (MAMI) and Jefferson Lab, thus adding a further piece of evidence towards an understanding of the ``proton radius puzzle". 

\section*{Acknowledgements}
We thank Achim Denig, Keith Griffioen, David Hornidge, 
Harald Merkel, Vladimir Pascalutsa, and Concettina Sfienti for useful discussions. 
This work was supported by the Deutsche Forschungsgemeinschaft (DFG) 
in part through the Collaborative Research Center [The Low-Energy Frontier of the Standard Model (SFB 1044)], and in part through the Cluster of Excellence [Precision Physics, Fundamental
Interactions and Structure of Matter (PRISMA)].

\end{document}